\documentclass{jfm}
\usepackage{subcaption}
\usepackage{graphicx}
\usepackage{newtxtext}
\usepackage{newtxmath}
\usepackage{natbib}
\usepackage[dvipsnames]{xcolor}
\usepackage{hyperref}
\hypersetup{
    colorlinks = true,
    urlcolor   = blue,
    citecolor  = black,
}

\newcommand{\RomanNumeralCaps}[1]
\linenumbers

\usepackage{silence}
\title{Airfoil tonal noise reduction by roughness elements \\ Part I -- Experimental investigation}

\author{El\'ias Alva\aff{1}
  \corresp{\email{eliasrean@ita.br}},
  Zhenyang Yuan\aff{2},
  Tiago B. Ara\'ujo\aff{1},
  Filipe R. do Amaral\aff{3},
  Ardeshir Hanifi\aff{2},
 \and Andr\'e V.G. Cavalieri\aff{1}}

\affiliation{\aff{1}Divis\~{a}o de Engenharia Aeron\'autica, Instituto Tecnol\'ogico de Aeron\'autica,  S\~{a}o Jos\'{e} dos Campos, SP, 12228-900, Brazil
\aff{2}FLOW Turbulence Lab., Department of Engineering Mechanics, KTH Royal Institute of Technology, Stockholm, Sweden
\aff{3}Institut Pprime, CNRS–Universit\'e de Poitiers–ENSMA, 86000 Poitiers, France}

\begin{document}

\maketitle

\begin{abstract}

Laminar separation bubbles around airfoils lead to the growth of instability waves, which enhances to acoustic scattering at the trailing-edge, forming a feedback loop that produces to tonal noise. To reduce the trailing-edge tonal noise, an array of roughness elements was used over a NACA0012 airfoil, at low angles of attack and moderate Reynolds number. Aeroacoustics and flow visualization experiments were performed for four configurations: a baseline smooth airfoil, two configurations with roughness elements on either airfoil surface (pressure side or suction side), and a fourth configuration with roughness elements at both airfoil surfaces. The roughness elements are made up of a row of spanwise periodically spaced cylinders, which were placed close to the mid-chord position. It is expected the proposed array of roughness elements stabilizes the Kelvin-Helmholtz instability mechanism that is present on the separated shear layer through the induction of streak structures. The separation bubble is assessed using oil flow visualizations, and the cases with roughness elements show the generation of streaks downstream of the array cylinders, disrupting the separation bubble. Acoustic measurements results show a decrease, and in some cases the total suppression, of the tonal noise at Reynolds numbers ranging from 0.6$\times$10$^5$ to 2.5$\times$10$^5$, and angles of attack ranging from 0 to 4 degrees.

\end{abstract}
\section{Introduction}
\label{Introduction}

New technological developments in urban mobility, as well as aviation and wind energy, come at the cost of new engineering challenges in recent years. A well-known problem with these applications is the noise generated by wings and blades. An undesirable effects is the trailing-edge tonal noise produced by the boundary layers undergoing transition to turbulence. This noise problem has lead to several studies over the last few decades \citep{paterson1973vortex,tam1974discrete,arbey1983noise,brooks1989airfoil,mcalpine1999generation,desquesnes2007numerical,probsting2015laminar,yakhina2020experimental}. The presence of this unwanted phenomenon motivates the development of novel solutions to reduce tonal noise in airfoils, thereby contributing to the reduction of noise pollution.

Tonal noise in airfoils has been widely studied and recognized in regimes of moderate chord-based Reynolds number $Re_c$ of order $10^5$, and low angles of attack \citep{paterson1973vortex,arbey1983noise}. These studies show a relationship between the free-stream velocity $U_{\infty}$ and the frequency of the strongest tone, which scales with the power of $U^{1.5}_{\infty}$. Besides, such primary tone, measurements recognize secondary tones whose frequency scale with $U^{0.85}_{\infty}$. Later studies indicate tonal noise occurs due to a feedback loop mechanism \citep{tam1974discrete,arbey1983noise,mcalpine1999generation}. The feedback loop begins with instabilities in the boundary layer at the airfoil pressure side, linked to Tollmien-Schlichting instability waves (T-S), in laminar boundary layers. On the other hand, the laminar separation bubble and the resulting separated shear layer can support Kelvin-Helmholtz (K-H) instabilities \citep{atassi1984feedback}. A recent numerical study \citep{nguyen2021numerical} suggests the feedback loop mechanism is related to both T-S waves in the attached zones (slowly-growing instabilities) and switches to K-H waves in the separated regions (fast-growing instabilities). Such waves grow until they reach the airfoil trailing-edge, generating acoustic waves via edge scattering \citep{williams1970aerodynamic}, which, in turn, travel upstream coupling to the instabilities waves, thus closing the loop. These mechanisms are related to instability and acoustic waves, and may thus be modelled using linear theory \citep{de2014global,demange2024resolvent}. A study by \citealt{desquesnes2007numerical} suggests vortex shedding from both pressure and suction sides also generates a feedback loop, contributing to tonal noise. Secondary tones emerge as a secondary bifurcation of this primary feedback loop, creating additional peak frequencies associated to a torus \citep{sano2023emergence}.

A necessary condition for tonal noise is the occurrence of a separation bubble near the trailing-edge \citep{mcalpine1999generation}. The regime that generates the tonal noise has been associated with the suction side for low Reynolds numbers and switches to the pressure side for moderate Reynolds numbers \citep{probsting2015regimes,probsting2015laminar}. A recent study of tonal noise with large-eddy simulation (LES) \citep{ricciardi2022switch} shows that, for low Reynolds numbers, the emission of tonal noise by the instability events on the suction side is predominant over the events on pressure side. Conversely, for high Reynolds numbers, the emission of tonal noise by events on the pressure side is predominant over the events on the suction side.

Different solutions were proposed to mitigate the trailing-edge tonal noise, as leading-edge and trailing-edge serrations \citep{roger2013reduction,chen2016experimental,oerlemans2001experimental,vathylakis2015poro}. These techniques were based on geometrical modifications, changing either leading or trailing-edge in order to affect overall aerodynamics \citep{serson2017direct} or trailing-edge scattering \citep{howe1991aerodynamic}. The possibility to directly modify the instability mechanisms of the laminar separation bubbles remains unexplored. This study proposes a novel technique for tonal noise attenuation through a passive control (roughness elements) aiming at generating streaks. Streaks are streamwise elongated structures \citep{andersson1999optimal}, emerging due to the lift-up mechanism \citep{brandt2014lift} and have spatial transient growth, as they decay by viscous effects, but streamwise vortices lead to the formation of streaks with large downstream extents \citep{andersson2001breakdown}. \citet{fransson2004experimental} performed experiments, with the generation of streaks by an array of cylinders on a flat plate. Such streaks have an stabilizing effect on the T-S waves \citep{fransson2005experimental}. Streaks also have an stabilizing effect on Kelvin-Helmholtz instabilities \citep{marant2018influence}. Such structures also have the potential to decrease the separation extent of the bubble (\citealt{karp2020optimal}). It has also been suggested that streaks have stabilizing properties for K-H instabilities \citep{marant2018influence}, which are the dominant instability at the separation bubble. Studies by \citet{sano2019trailing} indicated the presence of streaks in the boundary layer near the trailing-edge should have a negligible contribution to the radiated sound, thus suggesting the induction of streaks around the airfoil should not lead to a significant noise increase.

These properties motivated the study of the effect of roughness elements on the tonal noise of a NACA0012 airfoil. Following the design of previous studies \citep{fransson2004experimental,pujals2010drag}, we have excited steady streaks in the boundary layer in an attempt to reduce, or even suppress, the tonal noise of a NACA0012 airfoil. Streaks were generated by a row of spanwise periodically spaced cylindrical roughness elements close to the mid-chord position. Acoustic measurements show how tonal noise is affected by the roughness elements. We designed the roughness elements for a baseline case with $Re_c = 1\times10^5$ and zero degrees angle of attack, but the robustness of the proposed mechanism is assessed by changing the Reynolds number, angle of attack, and the airfoil side of the roughness elements in an extensive experimental campaign. A companion work \cite{Yuan2024airfoil} explores the baseline case in more detail using large-eddy simulations. 

This paper is structured as follows. In \S \ref{exp_setup}, we provide the experimental setup, measurement procedures and the design of roughness elements, whereas in \S \ref{Results} we present the results of surface pressure, oil flow visualizations, and aeroacoustic measurements for the airfoil without and with roughness elements. Finally, \S \ref{conclusions} completes the work with conclusions.
\section{Experimental setup and instrumentation}
\label{exp_setup}

All experiments presented here were carried out in an open circuit wind tunnel at the Instituto Tecnológico de Aeronáutica, which has a closed test section measuring 1.0 $\times$ 1.2 $\times$ 4.0 m$^3$, with a turbulence intensity in the test section of approximately $0.5$ \% \citep{assato2004research}. A NACA0012 airfoil spanning the test section, with 100 mm nominal chord ($c$), was used in this study. In order to ensure a fixed trailing-edge geometry for companion large-eddy simulations \cite{Yuan2024airfoil}, the airfoil trailing-edge is rounded with a curvature radius of $r_{te}/c = 0.4\%$ as in previous studies  \citep{ricciardi2022switch,demange2023experimental}, leading to a chord of $c^*$ = 96.62 $\pm$ 0.28 mm, as displayed in \ref{fig:naca0012_roughnesselements}. The model was vertically installed in the center of the test section. At the bottom, the model was fixed by a single strut which allows the use an external turntable. This mechanism enables the rotation of the model to easily adjust its angle of attack ($\alpha$). At the top, the airfoil was fixed by a bearing.

For the measurement of free stream dynamic pressure and the surface pressure around the airfoil, we used pressure transducers calibrated with a Betz manometer. The airfoil has 25 pressure taps, which were connected through a tubing system to two ESP-32HD scanning valves outside the wind tunnel. 
Both ESP were  connected to a National Instruments acquisition system, and the pressure acquisitions were synchronized for each measurement. 

The pressure coefficient $C_p$ was obtained using the pressure taps on the airfoil. The sampling rate for the pressure taps was 2000 samples per second, and 1000 samples were taken. The reference pressure for all taps was the static pressure at the inlet of the test section. The alignment of the airfoil was verified by comparing the measured pressure distribution at the two sides of the airfoil, and also by a comparison with reference numerical results obtained using XFOIL \citep{drela1989xfoil}.

\subsection{Design of roughness elements}

The roughness elements were designed following the approach of \citet{fransson2004experimental}. In order to generate streaks, we place cylinders on the surface of the airfoil. Since streaks have a stabilizing effects on T-S waves and K-H waves \citep{cossu2002stabilization,marant2018influence}, we expect that streaks induced by the cylinders disrupt the separation bubble on the airfoil surface, affecting significantly the feedback loop of tonal airfoil noise. In the study of \citet{fransson2004experimental} the relevant parameters were the local boundary layer thickness $\delta$, the height ($k$), the diameter ($d$), and the distance between each roughness elements ($\Delta z$), and the distance between the roughness elements and the leading-edge ($x$).
    \begin{figure}
        \centering
        \includegraphics[width=13cm]{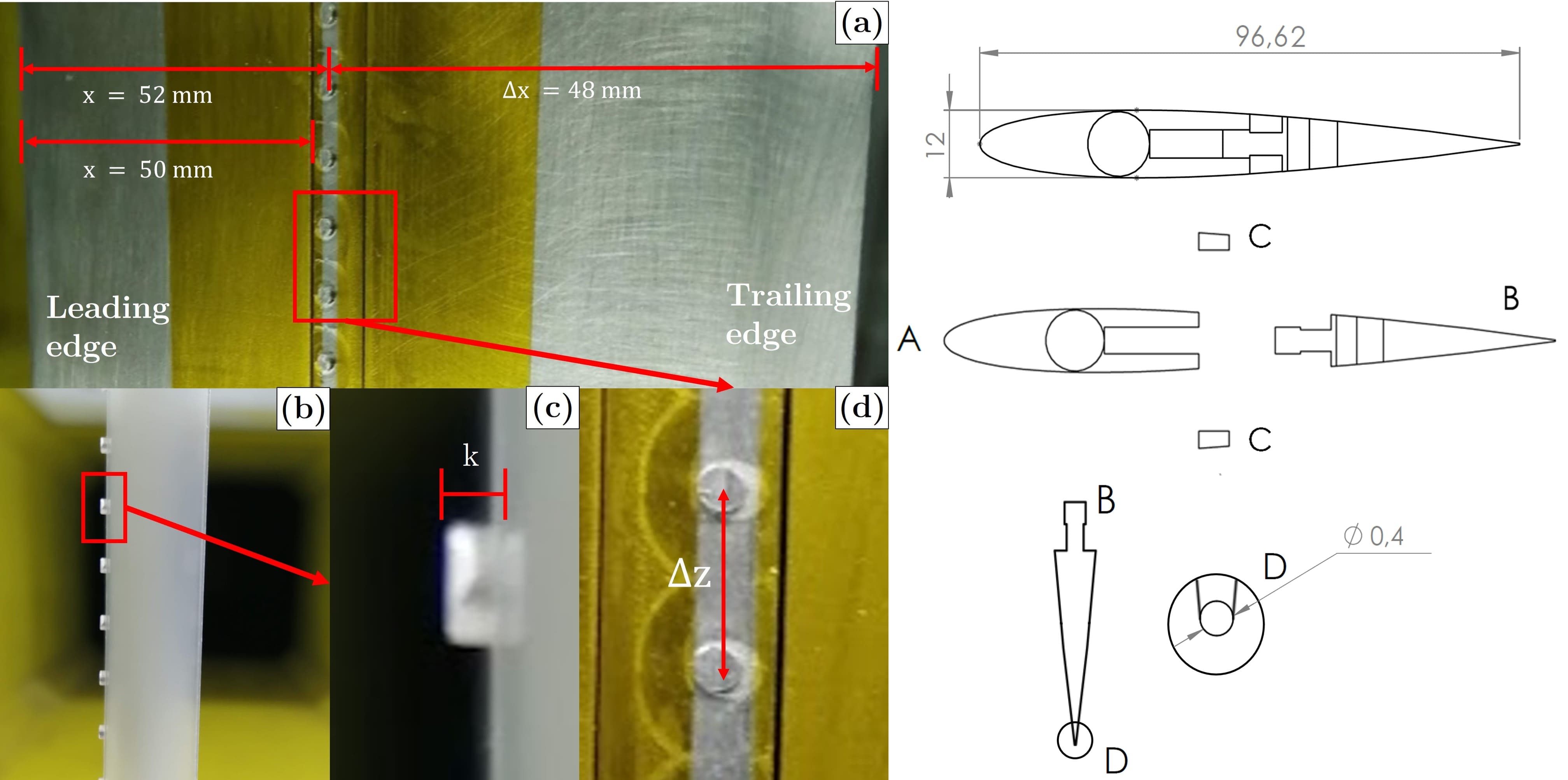}
        \caption{On the left, details of roughness elements used on the airfoil. The position of the roughness elements from the leading-edge in (a), the distance between each cylinder in (d), the height of the cylinders in (b) and (c). On the right, details of the NACA0012 airfoil with interchangeable strips used.}
        \label{fig:naca0012_roughnesselements}
    \end{figure}

    \begin{table}
        \begin{center}
        \begin{tabular}{lcccccccc}
             Configuration & $\delta$, mm& $x$, mm & $\Delta z$, mm & $k/\delta$ & $Re_{kk}$ & $\Delta z/\delta$ & $\Delta z/d$ & $d/k$ \\[3pt]
             \citeauthor{fransson2004experimental} & 0.27 & 40 & 8 & 2.84 & 340 & 29.06 & 4 & 2.56\\
             Present design & 0.23 & 55.4 & 6 & 2.39 & 336 & 26.08 & 4 & 2.7\\
        \end{tabular}
        \caption{\label{tab:roughness_parameters} Comparison of roughness parameters between \citet{fransson2004experimental} and the present design. $\delta$ is the local boundary-layer length scale, $x$ is the distance between the cylinders and the airfoil leading-edge (flat plate leading-edge in Fransson's approach), $\Delta z$ is the distance between cylinders, $k$ is the height of the cylinders, $d$ is the diameter of the cylinders, and $Re_{kk}$ is the Reynolds number measured at the height of the cylinder.}
        \end{center}
    \end{table}
For this study, the elements were designed for a Reynolds number of $Re_c$ = 1$\times$10$^5$, considering, for simplicity, a flat-plate boundary layer with a free-stream velocity $U_{\infty}$ = 15 m/s. The flat plate boundary layer amounts to neglecting the pressure gradients around the airfoil, which are small for zero and low angle of attack. We will see that streaks are induced for a range of angles of attack and Reynolds numbers, showing that this simplification does not compromise the proposed mechanism. We use similar parameters of \citet{fransson2004experimental} to design the roughness elements. The ratio is set as $\Delta z / \delta = 4$, and the location of roughness element was chosen as $x = 55.4$ mm, such that streaks could reach its maximum amplitude near the trailing-edge location. The value of other parameters were established using as reference the experiments of \citet{fransson2004experimental}, and considering geometric and manufacturing limitations. Using a Blasius profile, we have the local boundary layer thickness $ \delta= \sqrt{{x\nu}/{U_{\infty}}}= 0.23$ mm at the position $x$ of the roughness elements.  Table \ref{tab:roughness_parameters} shows the comparisons between \citeauthor{fransson2004experimental} and the present design. Figure \ref{fig:naca0012_roughnesselements} shows the roughness elements used on the NACA0012 airfoil. As shown in the figure, the roughness elements are part of an interchangeable strip, whose edges are covered with thin tape (shown in yellow in \ref{fig:naca0012_roughnesselements}, and also shown in figure \ref{fig:painted_area}) so as to avoid the appearance of gaps. The baseline smooth airfoil is obtained by using a smooth strip instead of the one with roughness elements. 

    \begin{figure}
    \centering
    \includegraphics[width=8cm]{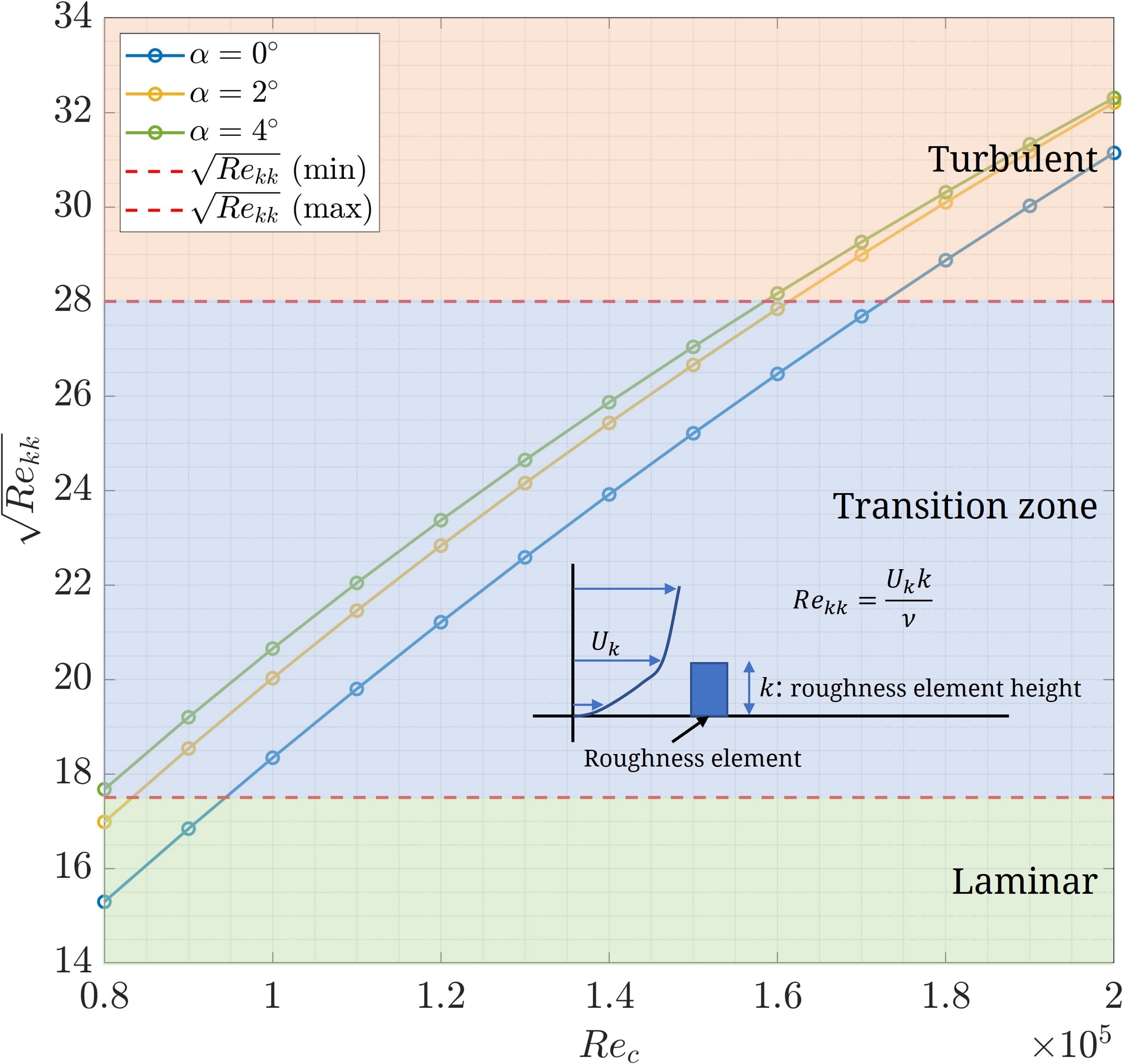}
    \caption{Estimated values of square root of $Re_{kk}$ in relation of $Re_c$ for different cases of $\alpha$ of this study.}
    \label{fig:rekk_alfa}
    \end{figure}

The roughness elements may have a stabilizing effect on Tollmien-Schlichting or Kelvin-Helmholtz waves by inducing streaks, but may also directly lead to transition to turbulence, which is already known to suppress tonal noise \citep{probsting2015regimes}, albeit with an expected drag penalty. \citeauthor{von1961effect} (\citeyear{von1961effect}) studied the effects of roughness elements on the boundary layer and reported the dimensions of the roughness elements leading to a transition to turbulence. The study evaluates the relation between the diameter ($d$) and the height ($k$) of the roughness elements as a function of $\sqrt{Re_{kk}} = \sqrt{U_k k /\nu}$, where $U_k$ is the velocity of non-disturbed flow at the height $k$, and $\nu$ is the kinematic viscosity.
The experimental study of \citeauthor{von1961effect} (\citeyear{von1961effect}) suggests there are some values of $\sqrt{Re_{kk}}$ where the transition to turbulence is effective, then the roughness elements could directly trip boundary layer (as tripping devices would do) instead of generating streaks. The present roughness design leads to $Re_{kk}= 336$, close to the experiments by \citet{fransson2004experimental}, who used $Re_{kk} = 340$. For the aspect ratio of our roughness elements, the value of $d/k = 2.7$ leads to initial transitional behavior for $\sqrt{Re_{kk}} = 17.5$. Figure \ref{fig:rekk_alfa} show the values of $\sqrt{Re_{kk}}$ for different $\alpha$. Green area represents the laminar values, blue area between dashed red lines shows the minimum and the maximum $\sqrt{Re_{kk}}$ that lead to transition to turbulence \citep{von1961effect}, and the turbulent area is shown in orange, taken from \citet{de2022transition}. For low Reynolds numbers $Re_c \leq 1\times 10^5$ and $\alpha = 0$ degrees, our values are outside of the range that leads to transition, and for $Re_c > 1.7 \times 10^5$ the experimental values in the literature suggest the roughness elements will induce a transition to turbulence in our experiments. Here, we used a $\sqrt{Re_{kk}} = 18.33$ for a $Re_c = 1 \times 10^5$, which is a value very close to the beginning of the transition for $\alpha = 0$ degrees; thus we do not expect to have direct transition to turbulence for the lower $Re_c$ which we also address in this study. On the other hand, for higher $Re_c$, a transition to turbulence may occur downstream. It is nonetheless interesting to evaluate the generation of streaks and how airfoil tonal noise is affected in these cases.

\subsection{Acoustic measurements}

\begin{figure}
        \centering
        \includegraphics[width=8cm]{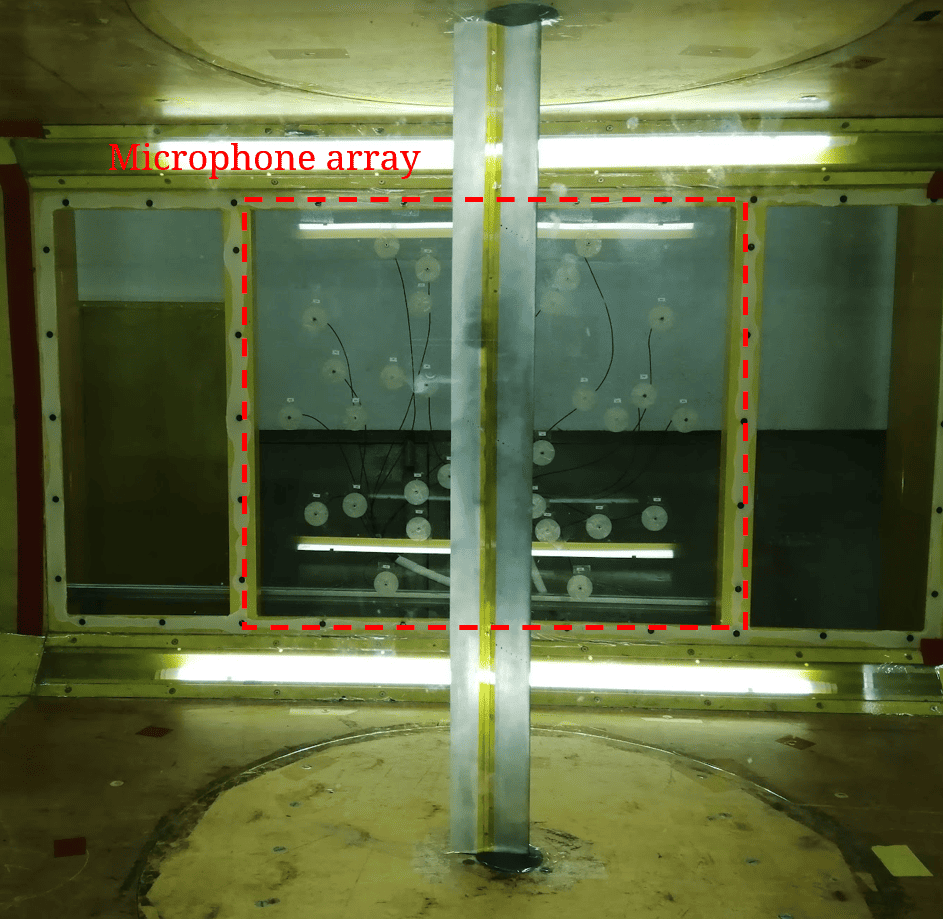}
        \caption{Experimental setup. The microphone array in the wall of the test section (second plane), parallel to the model (first plane).}
        \label{fig:wind_tunnel}
    \end{figure}

For the aeroacoustic measurements, we used an array with 34 PCB Piezotronics 130 Series microphones (figure \ref{fig:wind_tunnel}) connected to a Bruel \& Kjaer type 3560 acquisition system. The microphone array was placed on the wind tunnel wall, at a distance of 645 mm perpendicular to the chord. The microphone array design was based on the work by \citet{amaral2018design}. The microphones were calibrated with a pistonphone at an inlet pressure of 1 Pa, and 10 s of time acquisition. The parametric space addressed in the present study is given in table \ref{tab:aeroacoustic_parameters}. These parameters are close to the experiments performed by \citet{probsting2015regimes}, as the objective was to reduce tonal noise.
    \begin{table}
        \begin{center}
        \begin{tabular}{lccc}
             Case & Notation & $\alpha$ & Reynolds number \\[3pt]
             1 & Smooth surface on both sides & $0^{\circ}-4^{\circ}$ $\pm$ 0.1$^{\circ}$ & 0.6$\times10^5$ - 2.5$\times10^5$\\
             2 & Roughness elements on suctions side (S.S.) & $0^{\circ}-4^{\circ}$ $\pm$ 0.1$^{\circ}$ & 0.6$\times10^5$ - 2.5$\times10^5$\\
             3 & Roughness elements on pressure side (P.S.)& $0^{\circ}-4^{\circ}$ $\pm$ 0.1$^{\circ}$ & 0.6$\times10^5$ - 2.5$\times10^5$\\
             4 & Roughness elements on both sides & $0^{\circ}-4^{\circ}$ $\pm$ 0.1$^{\circ}$ & 0.6$\times10^5$ - 2.5$\times10^5$\\
        \end{tabular}
         \caption{\label{tab:aeroacoustic_parameters} Parametric space addressed in the aeroacoustic experiments.}
        \end{center}
    \end{table}

The acoustic data were post-processed with in-house conventional beamforming codes (\citealt{do2019improvements}) to improve the identification of potential noise sources, which is useful as the measurements are not taken in an anechoic environment. After a convergence study, the beamforming mesh domain dimensions were chosen as L$_x$ $\times$ L$_y$ = 600 $\times$ 1100 mm$^2$, with a region of interest of 200 $\times$ 800 mm$^2$ centered at the mid-span, mid-chord, and 645 mm from the array plane. To characterize the tonal noise as a function of the Reynolds number, experiments were carried out within an interval of free-stream velocity between 9 and 41 m/s with steps of approximately 0.5 m/s, which provides 66 velocities for a fine discretization of the trends of tonal noise with increasing Reynolds number. These velocities yield Reynolds numbers between $Re_c \approx 0.6 \times 10^5$ and $2.5 \times 10^5$ and Mach numbers in the $0.026 \leq M \leq 0.12$ range. In addition to these cases, the background noise of the wind tunnel was measured for the entire velocity range used in the present experiments.

The acquisition time for the acoustic measurements was set at 32 seconds with a sampling rate of 32.7 kHz. For spectral analysis, we used the Welch's method, with a block size $N_{fft} = 16384$ and a Hanning window with 75$\%$ of overlap. The parameters mentioned above give a frequency resolution of 2 Hz, corresponding to chord-based Strouhal numbers ranging from 1.12 to 38 in our experiments, and a number of blocks of 127. The PSD is evaluated in decibels by $10\times \log_{10}$($P$/($2\times10^{-5})^2)$, where $2 \times 10^{-5}$ Pa is the reference pressure and $P$ denotes the source auto-spectra values in Pa.

\subsection{Oil flow visualization}

\begin{figure}
        \centering
        \includegraphics[width=8cm]{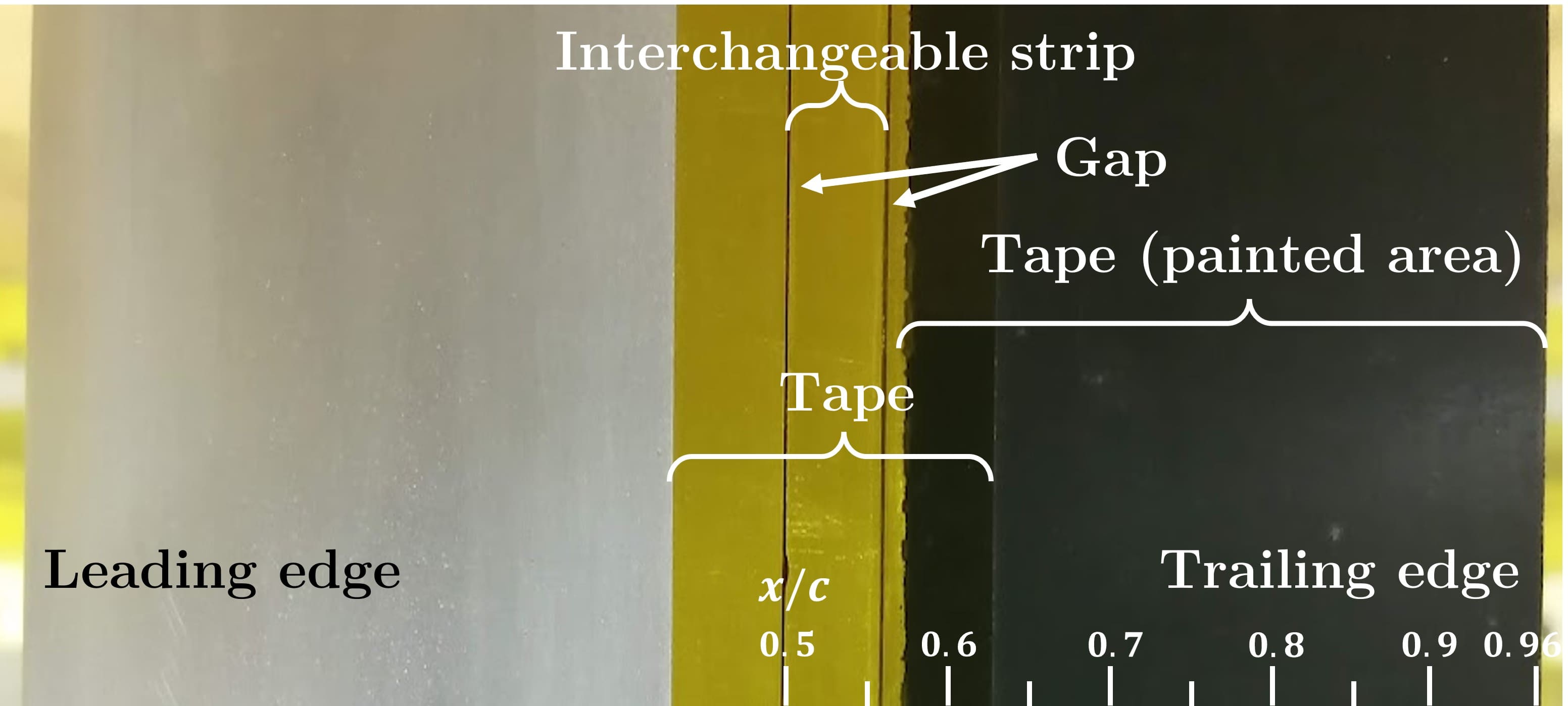}
        \caption{Setup for oil flow visualization, the flow is from left to right. The painted area in black is the region of interest.}
        \label{fig:painted_area}
    \end{figure}

Oil flow visualization was performed using a mixture of oleic acid, hydrated aluminum silicate (kaolin), and turpentine with a 1:16.7:68.9 volume ratio. The airfoil surface was prepared to highlight the marks left by the mixture, and thus we painted an area of the surface in black, where the results of the experiments were focused. Remaining openings and thin gaps between the interchangeable strip and the airfoil, where covered with tape, as shown in figure \ref{fig:painted_area}.
The painted area covers the model between $x/c = 0.5$ and the trailing-edge in order to observe the effects of the roughness elements on the surface flow. We apply the mixture with a roller to create a uniform boundary of mixture on the airfoil surface. With the airfoil mounted vertically, the tunnel flow needs to reach the desired experimental conditions as soon as possible to prevent the evaporation of the mixture and the downward movement of the mixture due to gravity. We performed experiments on the pressure and suction sides at angles of attack of 0, 1, 2, 3, and 4 degrees for cases with roughness elements and smooth surfaces. Reynolds numbers of $1\times10^5$, $1.2\times10^5$, and $1.5\times10^5$, where measured for the visualization experiments. The wind tunnel runs until the mixture is completely dry, achieving a steady state condition, and the pictures were taken at a fixed position with the wind tunnel off.
\section{Results}
\label{Results}
\subsection{Pressure coefficient}

\begin{figure}
    \begin{center}
    \includegraphics[width=13cm]{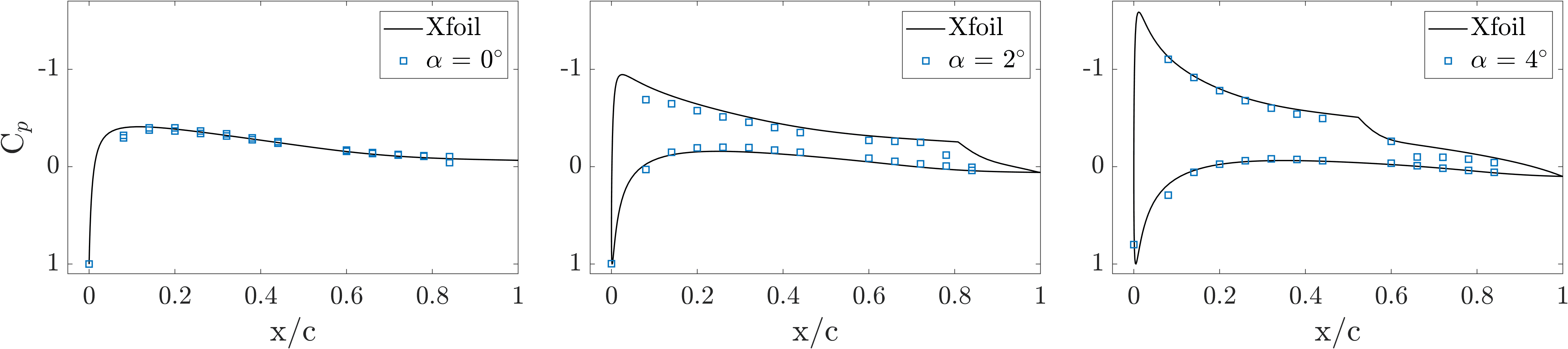}
    \caption{Pressure coefficient for NACA0012 at $Re_c$ = 1$\times$10$^5$ and $0$, $2$, and $4$ degrees of angle of attack. Blue squares denote a baseline case (smooth surface), and solid black lines indicates Xfoil results. }
    \label{fig:cp_re100}
    \end{center}
    \end{figure}

Figure \ref{fig:cp_re100} shows the pressure coefficient ($C_p$) for the baseline case (smooth surface) at three angles of attack ($\alpha = 0, 2$, and $4$ degrees). The results from the Xfoil panel method (\citealt{drela1989xfoil}), incorporating a boundary correction, natural transition, and a critical N-factor $N_{crit} = 7$, are also displayed.
For $\alpha = 0$ degrees, shown in figure \ref{fig:cp_re100} (a), the peak pressure is located at the leading-edge (stagnation pressure); in addition, the nearly identical pressures on the two airfoil surfaces ensures that the alignment of the model is accurate. The lack of significant differences between the effective angle of attack (induced by wind tunnel blockage) and the numerical one provided by Xfoil is probably due to a low blockage ratio of 1.2\% in the present setup, which can be considered as negligible.

\subsection{Flow visualization}
    \begin{figure}
    \begin{center}
    \includegraphics[width=13cm]{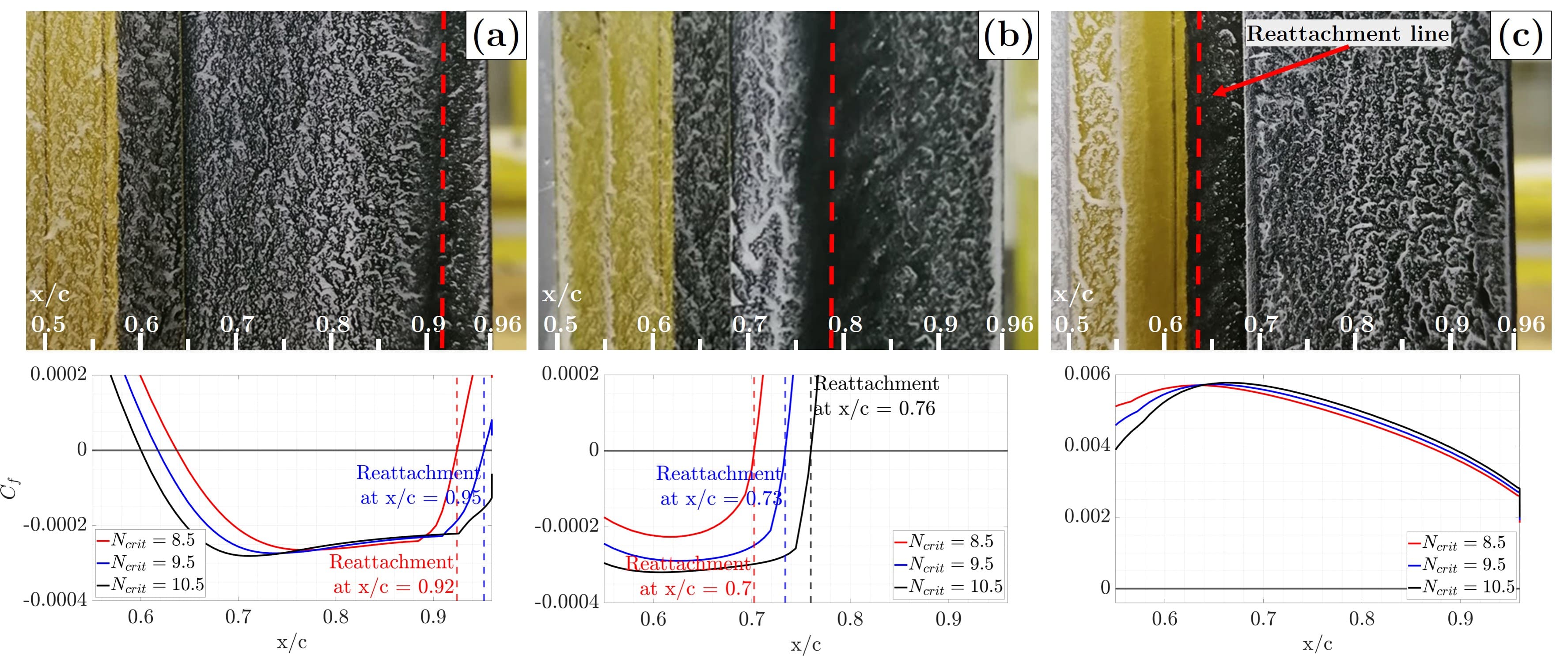}
    \caption{On the top, oil flow visualization on the suction side for smooth case and angles of attack of $0$, $2$ , and $4$ degrees, and $Re_c$ = 1.5$\times$10$^5$. Red dashed lines represent the reattachment position. On the bottom, numerical $C_f$ obtained by Xfoil for the same angles of attack and different $N_{crit}$. Dashed lines represent the reattachment position $C_f = 0$ for each numerical result. }
    \label{fig:reattachment_line}
    \end{center}
    \end{figure}

Figure \ref{fig:reattachment_line} shows the oil flow visualization and the skin friction coefficient evaluated using Xfoil software, for angles of attack of $0$, $2$, and $4$ degrees and Reynolds number equal to $1.5\times 10^5$. The flow goes from left to right for all cases. At the laminar separation bubble, it is observed an accumulation of oil due to the recirculation region, whereas the reattachment line is marked by the zone with an absence of oil on the surface, since the boundary layer after the reattachment has large friction coefficient and tends to clear the oil from the airfoil surface \citep{mcgranahan2003surface}. The position of the reattachment line may be checked with numerical results for the friction coefficient $C_f$, which should be zero at the reattachment zone. Figure \ref{fig:reattachment_line} shows the numerical results of $C_f$, considering free transition criterion in Xfoil with three different $N_{crit}$. As we can see, the values of $N_{crit}$ affect the position of the reattachment. However, for $\alpha = 0$, and $2$ degrees, the numerical results show reattachment points consistent with the experimental observation for $N_{crit} = 10.5$. On the other hand, for $\alpha = 4$ degrees predicts a laminar separation bubble and a reattachment point well upstream of the experimental observation. The reason for discrepancy is unclear.

The presence of streaks induced by the roughness elements on the airfoil can be verified with the oil flow visualization. In the experimental work of \cite{fransson2004experimental} the high speed streaks are generated downstream of the cylinders, and low speed streaks are generated between the cylinders.

    \begin{figure}
    \begin{center}
    \includegraphics[width=13cm]{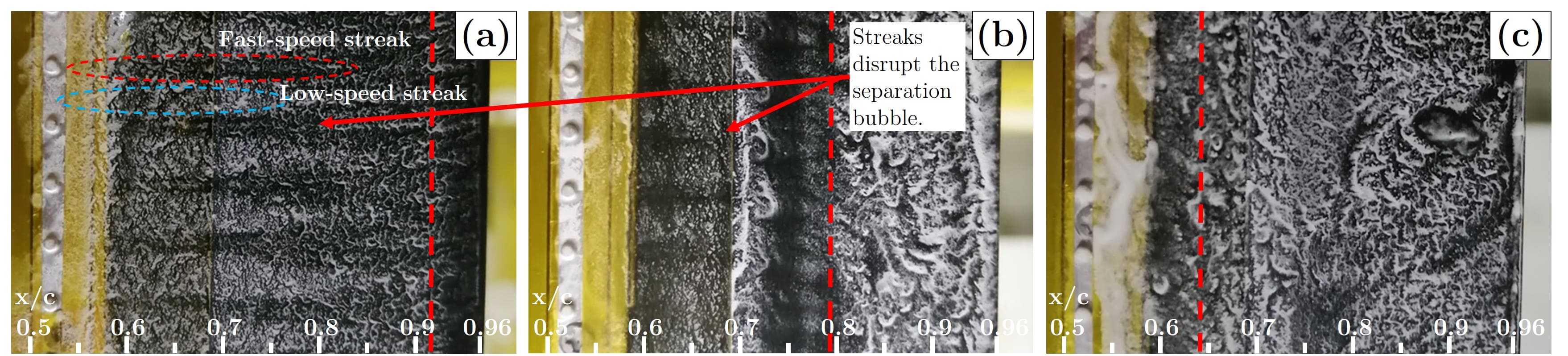}
    \caption{On the top, oil flow visualization of the suction side for roughness case and angles of attack of $0$, $2$, and $4$ degrees, and $Re_c$ = 1.5$\times$10$^5$, red dashed lines represent the reattachment position for smooth case. Flow goes from left to right.}
    \label{fig:oil_flow_roughness}
    \end{center}
    \end{figure}

    \begin{figure}
    \begin{center}
    \includegraphics[width=13cm]{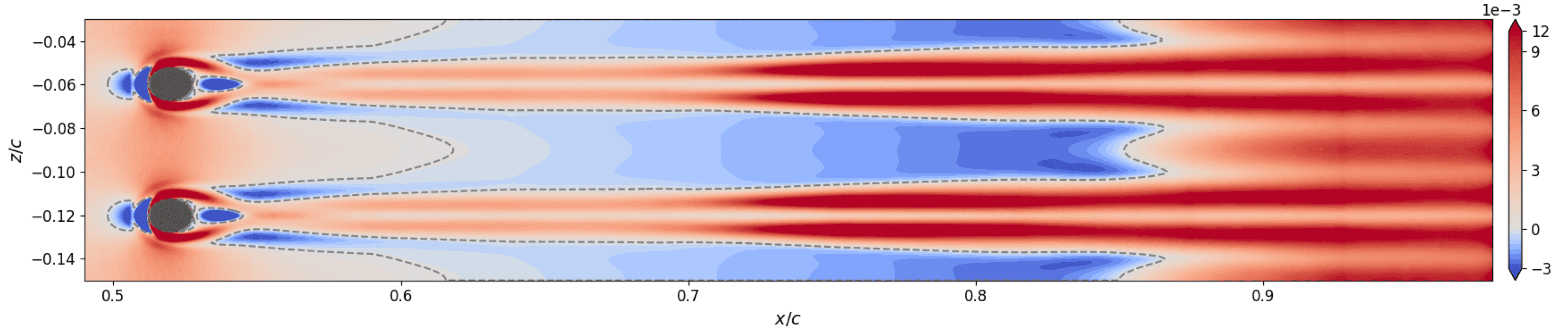}
    \caption{$C_f$ hot map of the mean flow field with roughness elements at $\alpha = 0$ degrees and $Re_c = 1 \times 10^5$ from our numerical study \cite{Yuan2024airfoil}. Red and blue represent positive and negative values, respectively.}
    \label{fig:cf_map}
    \end{center}
    \end{figure}

Figure \ref{fig:oil_flow_roughness} shows the oil flow visualization for the case with roughness elements on the suction side for $0$, $2$, and $4$ degrees of angle of attack and $Re_c = 1.5\times10^5$. For $\alpha = 0$ degrees, in figure \ref{fig:oil_flow_roughness} (a) the roughness elements generate high-speed streaks downstream of the cylinders featuring a weak tracing of oil on the surface. Results are consistent with higher friction coefficients in our numerical study \cite{Yuan2024airfoil}. The streaks develop in the region of the separation bubble as shown in figure \ref{fig:reattachment_line} (a); in this case, the roughness elements are located in the laminar boundary layer upstream of separation. Figure \ref{fig:oil_flow_roughness} (b) show the results for $\alpha = 2$ degrees. According the results shown in figure \ref{fig:oil_flow_roughness} (b), the roughness elements are already in a region of separated flow. The flow visualization shows some weak traces downstream of roughness elements, suggesting that weak high-speed streaks are developing on the surface. Both figures \ref{fig:oil_flow_roughness} (a) and (b) show streaks disrupting the separation bubble, which becomes three-dimensional, possibly with alternating regions of attached and separated flow; this is confirmed by our numerical simulations in the companion work by \cite{Yuan2024airfoil}. On the other hand, at $\alpha = 4$ degrees, figure \ref{fig:reattachment_line} (c), no significant effect of the roughness elements on the generation of streaks is observed, likely due to a turbulent boundary layer developing upstream of the roughness elements. Although streaks may be induced by roughness elements in turbulent boundary layers \citep{pujals2010drag}, the cylinders in the present experiments were designed considering a laminar boundary layer with zero pressure gradient, and are likely ineffective for turbulent boundary layers at higher angles of attack.

\begin{figure}
        \begin{center}
        \includegraphics[width=13cm]{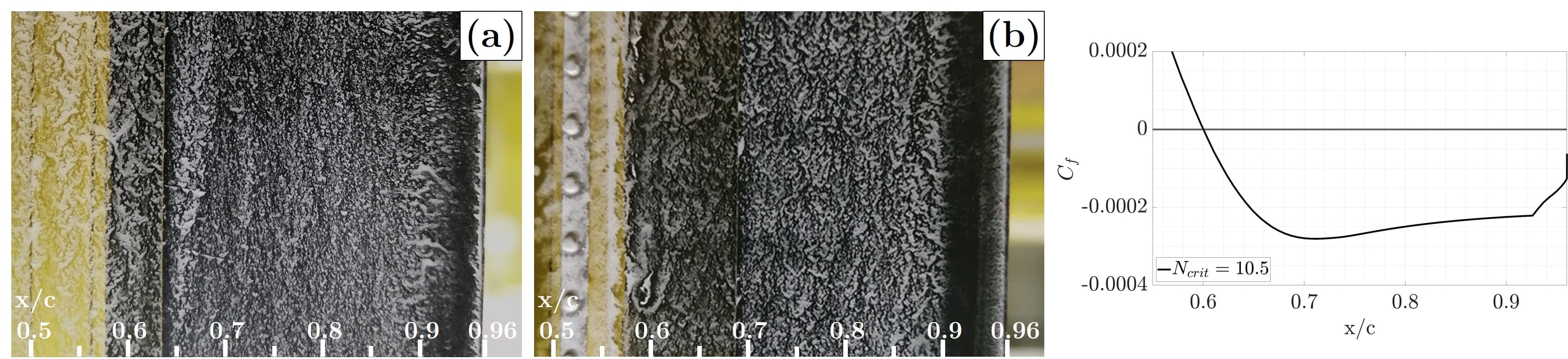}
        \caption{Oil flow visualization for smooth (a) and rough (b) cases at $\alpha = 0$ degrees and $Re_c = 1\times 10^5$. On the right, the skin friction coefficient for the same parameters.}
        \label{fig:oil_a0_re100}
        \end{center}
    \end{figure}

Figure \ref{fig:oil_a0_re100} shows results for Reynolds number of $1\times 10^5$ and $\alpha = 0$ degrees for both smooth and rough surfaces. At this lower Reynolds number, oil flow visualization becomes harder due to the lower wind tunnel velocities, but some characteristics of the higher $Re_c$ shown in figures \ref{fig:reattachment_line} and \ref{fig:oil_flow_roughness} remain. 
Figure \ref{fig:oil_a0_re100} (a) shows the smooth case, suggesting a reattachment region close to the trailing-edge, as the separation bubble develops upstream. The $C_f$ computed shows separation close to $x/c = 0.6$ up to the airfoil trailing-edge, consistent with the experimental visualization. Figure \ref{fig:oil_a0_re100} (b) shows the case with roughness elements. Some weak traces show the development of streaks downstream of the cylinders, and these are clear until $x/c = 0.8$. This indicates a three-dimensionalization, and possibly a disruption, of the separation bubble by the streaks induced by the roughness elements. 
    \begin{figure}
        \begin{center}
        \includegraphics[width=13cm]{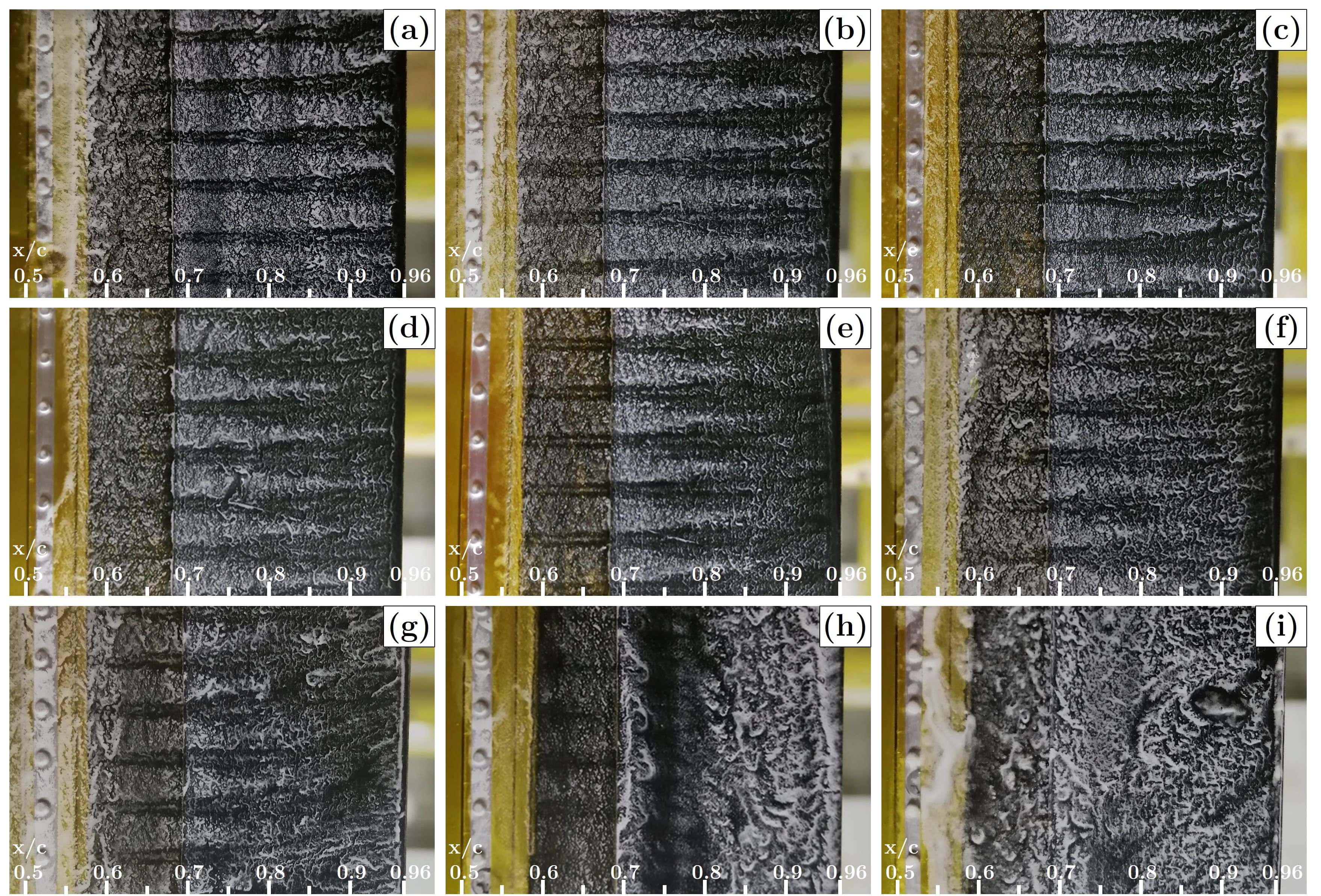}
        \caption{Oil flow visualization for roughness cases for the pressure side and, from frames (a) to (f), angles of attack $\alpha = 6$, $4$, $3$, $2$, $1$, and $0$ degrees, respectively. Frames (g) to (i) denote the and on suction side and angles of attack $\alpha = -1$, $-2$, and $-4$ degrees, respectively.}
        \label{fig:oil_ss_ps}
        \end{center}
    \end{figure}

Figure \ref{fig:oil_ss_ps} displays the evolution of the streaks as pressure gradient is modified by the variation of the angle of attack and for $Re = 1.5 \times 10^5$. Figures \ref{fig:oil_ss_ps} (a-f) represent the pressure side for angles of attack $\alpha = 6, 4, 3, 2, 1$, and $0$ degrees. For all these angles, we can observe the disruption of the separation bubble by the streaks; we note that for all of these cases the roughness elements are located upstream of the reattachment point. Moreover, the streaks are observed up to the trailing-edge region. On the other hand, figures \ref{fig:oil_ss_ps} (g-h) display the results for angles of attack $\alpha = 1$, and $2$ degrees on the suction side. Such angles of attack lead to a weaker effect of streaks on the separation bubble until $\alpha = 4$ degrees, figure \ref{fig:oil_ss_ps} (i), when the roughness elements no longer present visible effects on the generation of streaks, likely due to the introduction of roughness elements in a region of turbulent boundary layer, as discussed above.

\subsection{Acoustics}

\subsubsection{Acoustic spectra for representative Reynolds number}
\begin{figure}
        \begin{center}
        \includegraphics[width=13cm]{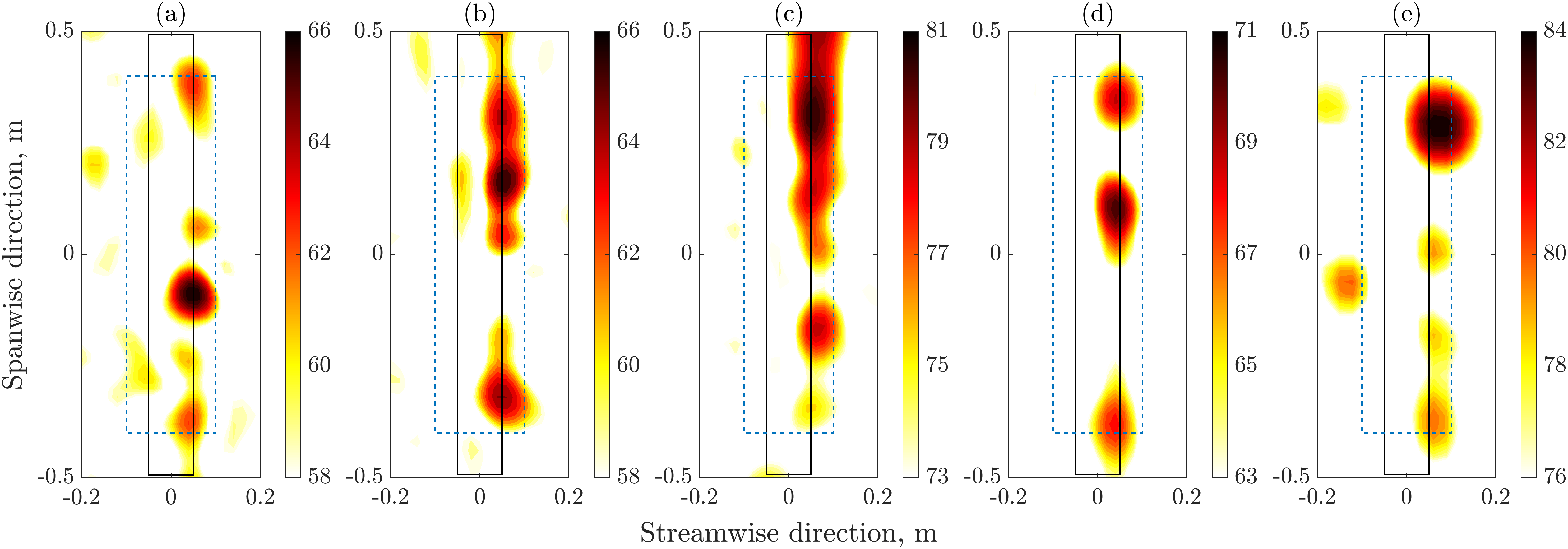}
        \caption{Conventional beamforming radiation maps  for the baseline case with a $Re_c = 2.5\times 10^5$ and contour plot in dB/St. (a) $\alpha = 0$ degrees and $St = 8.75$  , (b) $\alpha = 1$ degree and $St = 9.06$, (c) $\alpha = 2$ degrees and $St = 7.5$, (d) $\alpha = 3$ degrees and $St = 8.75$, and (e) $\alpha = 4$ degrees and $St = 6.88$. The free-stream velocity goes from left to right. In blue dashed lines the region of interest.}
        \label{fig:CBF_baseline}
        \end{center}
    \end{figure}

For acoustics results, we initially explore conventional beamforming (CBF), which is helpful to isolate sound sources emitted by the region of interest. This is especially important in the current experiments, since they are carried out in a hard-wall close test section. Figure \ref{fig:CBF_baseline} displays the conventional CBF maps for the baseline case and angles of attack between $0$ and $4$ degrees at Reynolds number $Re_c = 2.5 \times 10^5$. Figures (a) to (e) show the maps for Strouhal number (based on the chord $St = f c/U_{\infty}$) $St = 8.75, 9.06, 7.5, 8.75$, and $6.88$ respectively. This Reynolds number and Strouhal numbers show satisfactory results for CBF, with peak sound radiation from the trailing-edge region, as expected. The NACA0012 airfoil is presented as a rectangle with spanwise and streamwise extent of $10 c$, and $c$, respectively. The free stream velocity goes from left to right. The spots show the noise sources on the chord plane and are located at the trailing-edge of the airfoil for all cases, which ensures the consistency of present results.

    \begin{figure}
        \begin{center}
        \includegraphics[width=13cm]{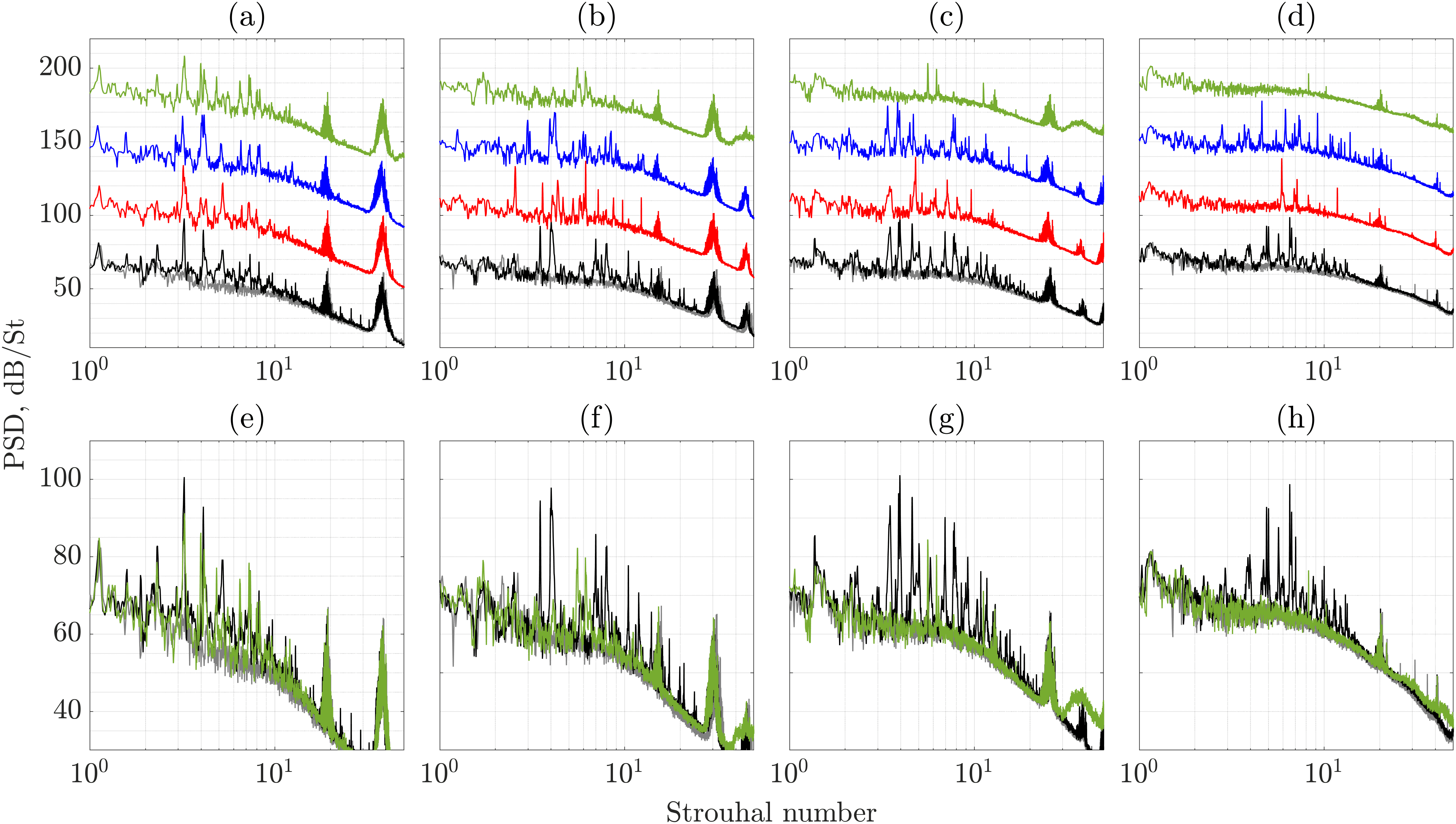}
        \caption{Narrowband PSD for $\alpha = 0$ degrees and $ Re_{c} = 0.8\times 10^5$ (a,e), $1\times 10^5$ (b,f), $Re_{c} = 1.2\times 10^5$ (c,g), $Re_{c} = 1.5\times 10^5$ (d,h). Frames (a-d) display the tunnel background noise in grey lines, smooth surface in black lines, roughness on suction side (S.S.) in red lines, roughness on pressure side (P.S.) in blue lines, and roughness on both sides in green lines. Each spectrum has an offset of 40 dB/St for visualization effect. Frames (e-h) display only the tunnel background noise (gray lines), the smooth surface(black lines), and roughness on both sides (green lines)}
        \label{fig:spectrum_a0}
        \end{center}
    \end{figure}

    \begin{figure}
        \begin{center}
        \includegraphics[width=13cm]{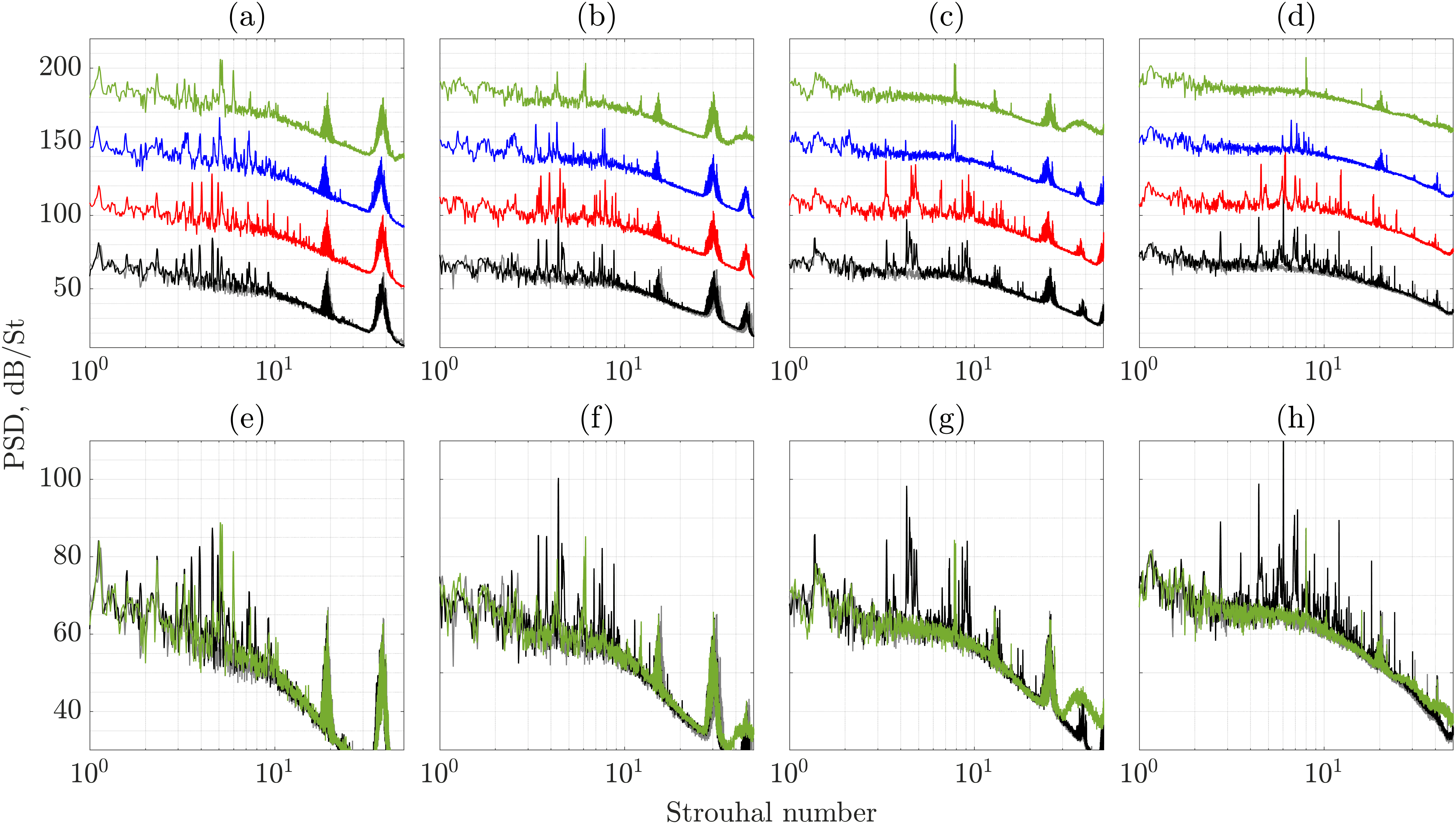}
        \caption{Narrowband PSD for $\alpha = 2$ degrees, and $ Re_{c} = 0.8\times 10^5$ (a,e), $1\times 10^5$ (b,f), $1.2\times 10^5$ (c,g), $1.5\times 10^5$ (d,h). Frames (a-d) display the tunnel background noise in grey lines, smooth surface in black lines, roughness on suction side (S.S.) in red lines, roughness on pressure side (P.S.) in blue lines, and roughness on both sides in green lines. Each spectrum has an offset of 40 dB/St for visualization effect. Frames (e-h) display only the tunnel background noise (gray lines), the smooth surface(black lines), and roughness on both sides (green lines).}
        \label{fig:spectrum_a2}
        \end{center}
    \end{figure}

    \begin{figure}
        \begin{center}
        \includegraphics[width=13cm]{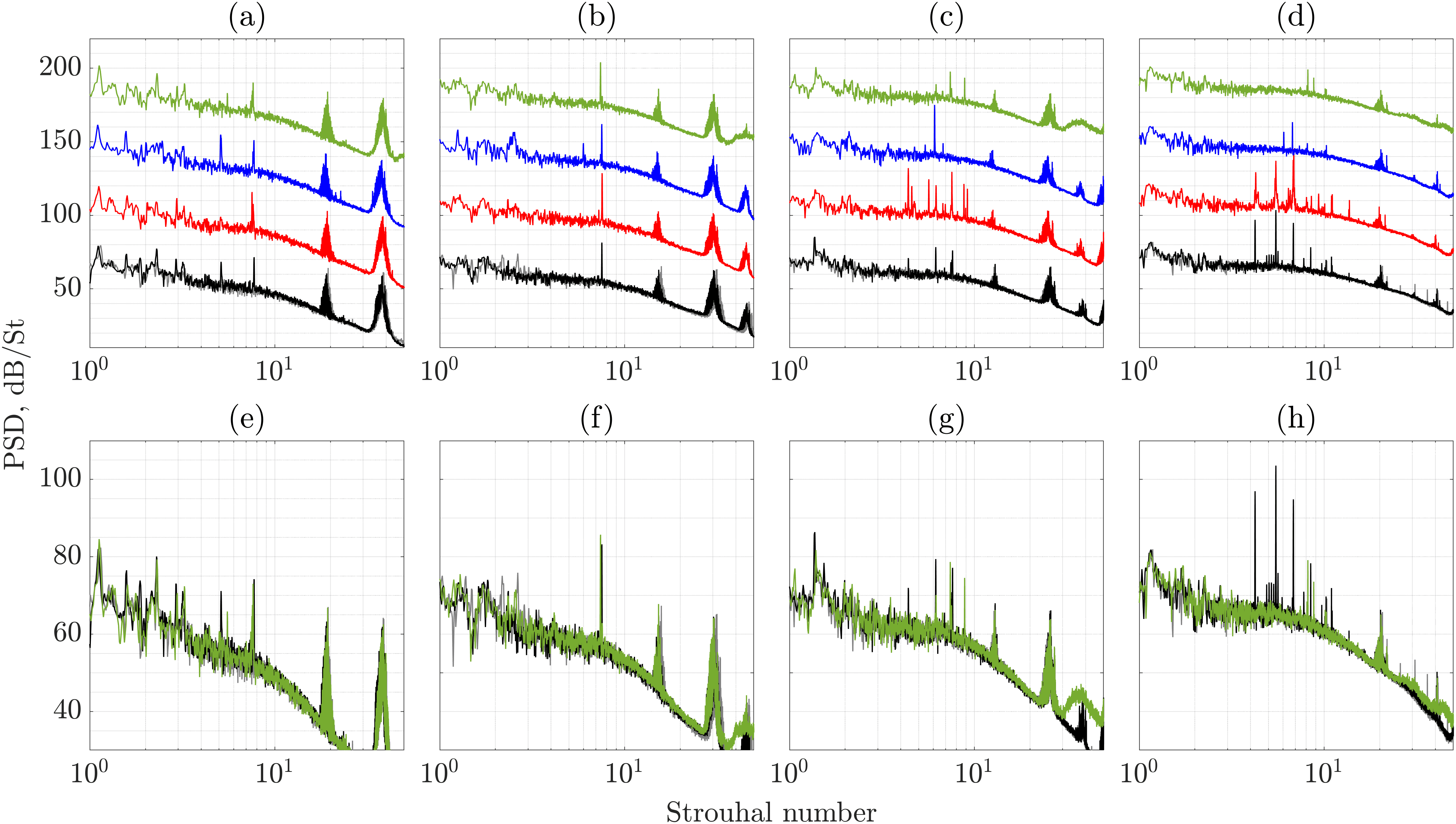}
        \caption{Narrowband PSD for $\alpha = 4$ degrees, and $ Re_{c} = 0.8\times 10^5$ (a,e), $1\times 10^5$ (b,f), $1.2\times 10^5$ (c,g), $1.5\times 10^5$ (d,h). Frames (a-d) display the tunnel background noise in grey lines, smooth surface in black lines, roughness on suction side (S.S.) in red lines, roughness on pressure side (P.S.) in blue lines, and roughness on both sides in green lines. Each spectrum has an offset of 40 dB/St for visualization effect. Frames (e-h) display only the tunnel background noise (gray lines), the smooth surface(black lines), and roughness on both sides (green lines).}
        \label{fig:spectrum_a4}
        \end{center}
    \end{figure}
Acoustic spectra for different cases with and without roughness elements are shown in figure \ref{fig:spectrum_a0}. The spectra are obtained after processing with CBF maps, retaining the sound generated by the region of interest as illustrated in figure \ref{fig:CBF_baseline}. We perform the acoustic experiments to evaluate the influence of streaks on the acoustic field. Also, the evaluation of the results at different angles of attack and Reynolds numbers will give us an overview of the robustness of the roughness elements.

Figure \ref{fig:spectrum_a0} shows the spectra for $\alpha = 0$ degrees, in frames (a,e) $Re_c = 0.8\times 10^5$, frames (b,f) $Re_c = 1\times 10^5$, frames (c,g) $Re_c = 1.2\times 10^5$, and frames (d,h) $Re_c = 1.5 \times 10^5$. The top frames show the four cases with an offset among the cases, i.e. smooth surface in black lines, roughness on suction side (S.S) in red lines, roughness on pressure side (P.S) in blue lines, and roughness in both sides in green lines. The bottom frames compare the smooth surface with the airfoil with roughness elements in both sides to show how much difference is obtained for the spectral peaks. For all frames, the tunnel background noise (obtained with CBF for the empty test section) is shown with grey lines. Since the airfoil is symmetric, the distinction between pressure and suction sides is arbitrary; here we follow the placement of roughness elements applied for positive angles of attack and investigate how roughness at a single side of the airfoil modifies sound radiation. In this case, the main tonal noise is measured at $St \approx 3.2, 4, 4$ and $6.5$ for $Re_c =$ 0.8, 1, 1.2, and 1.5 $\times 10^5$ respectively, with amplitudes more than 20 dB/St above the background noise. We can also notice the secondary tones, with amplitudes more than 10 dB/St above the background noise for $Re_c =$ 0.8, 1, 1.2, and 1.5 $\times 10^5$. The number of secondary peaks increases for $Re_c =$ 1, and 1.2 $\times 10^5$ and begins to decay for $Re_c = 1.5\times 10^5$, which is consistent with earlier experiments by \cite{probsting2015regimes}. The cases with roughness elements on either suction and pressure side lead to slight differences with respect to the baseline case, with a decrease of the number of secondary tones. Close to Strouhal numbers 19 and 38 at $Re_c = 0.8\times 10^5$, 15 and 30 at $Re_c = 1\times 10^5$, 13 and 26 at $Re_c = 1.2\times 10^5$, and 20 at $Re_c = 1.5\times 10^5$, the spectra display a set of peaks due to background noise. The results show that roughness elements only on one side have slight effects on tones for $Re_c = 0.8\times 10^5$. On the other hand, for $Re_c = 1$, and $1.2\times 10^5$ single-sided roughness elements reduce the primary peak and suppress some secondary peaks. Regarding the comparison between the smooth surface and the case with roughness on both sides, figures \ref{fig:spectrum_a0} (e-h), for all $Re_c$ the roughness elements on both sides (green lines) lead to tonal level decrease and, for some cases, suppression of the tonal noise. For $Re_c = 0.8\times 10^5$, figure \ref{fig:spectrum_a0} (e), the dominant tonal noise in the case with roughness on both sides presents a decrease of 9 dB/St; some other tones appear at $St =$ 4.8, 12, but weaker than the baseline case. Figures (f-h) display a clear suppression of the tonal noise. The case with roughness elements on both sides shows a suppression of primary tonal noise peaks and attenuation of secondary peaks for all cases.

Figure \ref{fig:spectrum_a2} shows the spectra for $\alpha = 2$ degrees, and $Re_c =$ 0.8, 1, 1.2, and 1.5 $\times 10^5$ in the same fashion as figure \ref{fig:spectrum_a0}. The smooth surface case (black solid lines) presents dominant tonal noise at St $\approx$ 4.6, 4.4, 4.3, and 6 for $Re_c = 0.8, 1, 1.2$, and $1.5 \times 10^5$ respectively, with amplitudes more than 15 dB/St above the background noise. The roughness elements on S.S. (red lines) show similar behavior to the baseline case, and some weaker tones appear for different Strouhal numbers. The roughness elements on P.S. (blue lines) lead to a decrease in the main tone, and some weaker tones were also suppressed for cases (b), (c), and (d) with respect to the background noise. When the roughness elements are on both sides (green lines) the effects on the tonal noise are stronger than in the cases with roughness elements, with most of the weaker tones being suppressed. Roughness elements are observed to reduce the tonal noise, but shifting the main tone frequency, suggesting a different noise suppression mechanism than tripping devices. Frames \ref{fig:spectrum_a2} (e-h) show the difference between the smooth case and the roughness on both sides, which completely suppress tonal noise \citep{probsting2015regimes}, likely with a corresponding drag penalty due to turbulent boundary layers.

Figure \ref{fig:spectrum_a4} displays the results for $\alpha = 4$ degrees, with the same Reynolds numbers of previous figures. The results for the smooth airfoil (black solid lines) present dominant tonal noise at St $\approx$ 7.7, 7.4, 6.1, and 5.5 for $Re_c = 0.8, 1, 1.2$, and $1.5 \times 10^5$ respectively. Only the cases with the $Re_c = 1.2$, and $1.5 \times 10^5$ present secondary and weaker tones; when the roughness elements are placed at the suction side, other weaker tones appear as well. The roughness cases do not show a great effect for low Reynolds numbers. However, when the roughness elements are on both sides (green lines) for a moderate Reynolds number, they suppress the main tonal noise and secondary tones above of the background noise. The same behavior appears when the roughness is located on the pressure side (blue lines). In frames \ref{fig:spectrum_a4} (e,g,h) roughness elements on both sides have a similar effect on tonal noise for $Re_c = 0.8$, $1$, and $1.5\times 10^5$, decreasing and sometimes suppressing tones. Conversely, for $Re_c = 1\times 10^5$, frame \ref{fig:spectrum_a4} (f), suggests an small increase of the amplitude of tonal noise, a situation that deserves further investigation.

\subsubsection{Trends of tonal noise with increasing Re}

The acoustic results are now presented in $St-Re$ contour plots for different configurations. The vertical axis represents the non-dimensional frequency based on the free-stream velocity and airfoil chord (Strouhal number) and the horizontal axis represents the chord-based Reynolds number. Each column of the plot represents the spectrum at a certain Reynolds number.

    \begin{figure}
        \begin{center}
        \includegraphics[height=6.5cm]{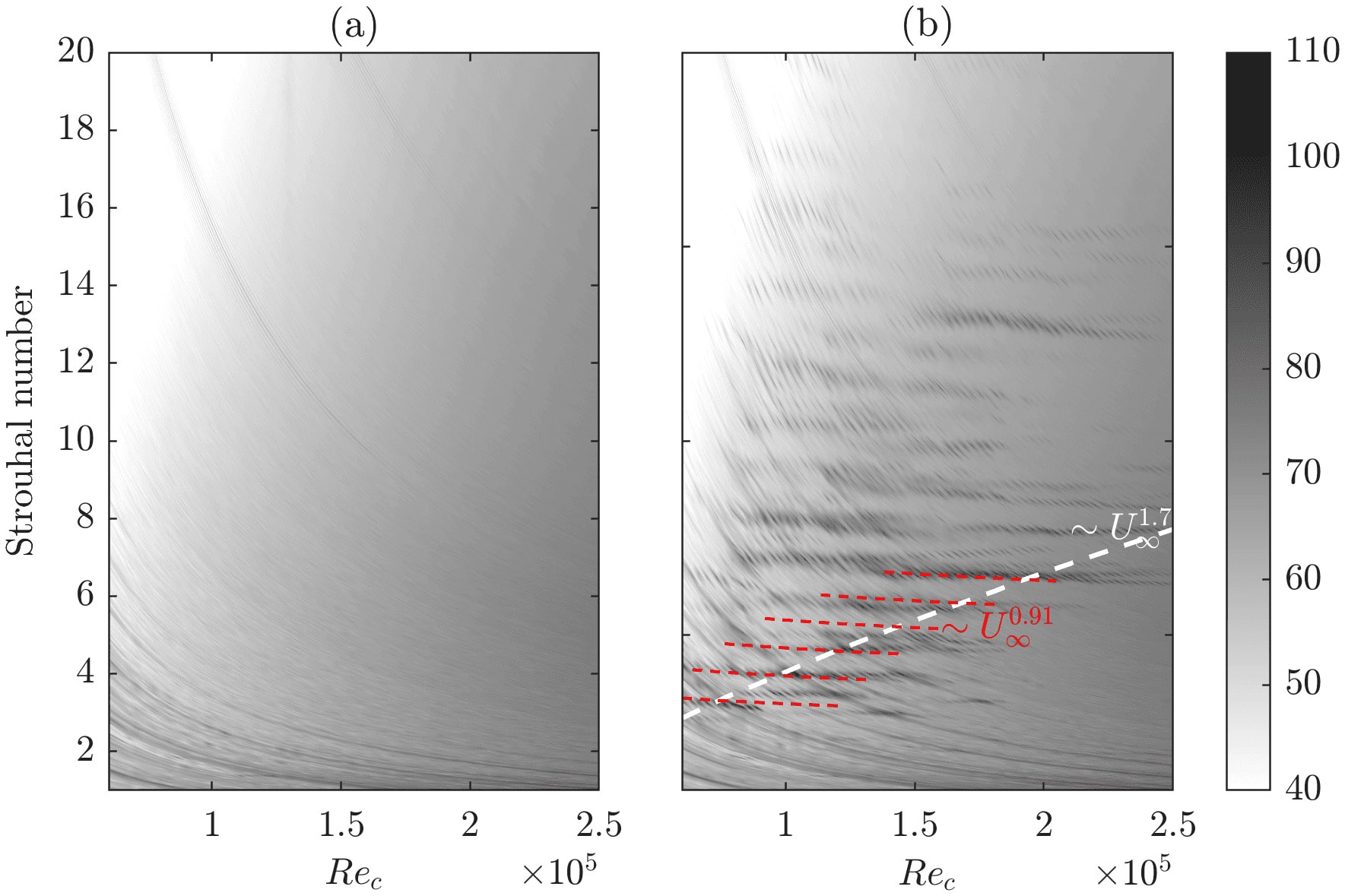}
        \caption{PSD (in dB/St) for background noise (a) and the smooth case at $\alpha$ = 0 degrees (b). White dashed line indicates the primary tone is proportional to $U_{\infty}^{1.7}$, and red lines denote the secondary tones are proportional to $U_{\infty}^{0.91}$.}
        \label{fig:map_BNa0}
        \end{center}
    \end{figure}
    
Figure \ref{fig:map_BNa0} shows the radiation map for (a) the tunnel background noise and (b) for $\alpha$ = 0 degrees smooth case. The baseline case displays several peaks of tonal noise in black spots for every Reynolds numbers, similar to that reported by \cite{paterson1973vortex}. Such peaks are significantly higher than the background noise. The Strouhal number of dominant peaks grows with increasing free-stream velocity. The main tonal noise peak follows the empirical formula 
    \begin{equation}
        f = 0.0075 U_{\infty }^{1.7} \frac{1}{(c \nu)^{0.5}},
        \label{eq:pri_tones}
    \end{equation}
with exponent 1.7, and constant 0.0075 differing slightly from what is shown in \cite{paterson1973vortex} for Reynolds number above $10^6$ (1.5 and 0.011, respectively). The dashed white line in figure \ref{fig:map_BNa0} (b) denotes the results of equation \ref{eq:pri_tones}. The main peaks for the baseline at 0 degrees angle of attack have consistent results with the predictions made by equation \ref{eq:pri_tones}. The frequencies of secondary tones are proportional to the free stream velocity by the empirical relation (\citealt{arbey1983noise})
    \begin{equation}
        f_n = \frac{K}{L_f} (n + 0.5 ) U_{\infty }^{0.91},
        \label{eq:sec_tones}
    \end{equation}
where $f_n$ are the discrete secondary frequencies, $K = 0.81$ is a proportionality constant and $L_f$ is the distance, in meters, between the maximum velocity point in the suction side and the trailing-edge, which was taken from Xfoil calculations. This relationship associates the feedback loop by the term $L_f$, with the tonal noise. The dashed red line in figure \ref{fig:map_BNa0} indicates the results for equation \ref{eq:sec_tones}, confirming that results are consistent with this empirical relation.

    \begin{figure}
        \begin{center}
        \includegraphics[width=13cm]{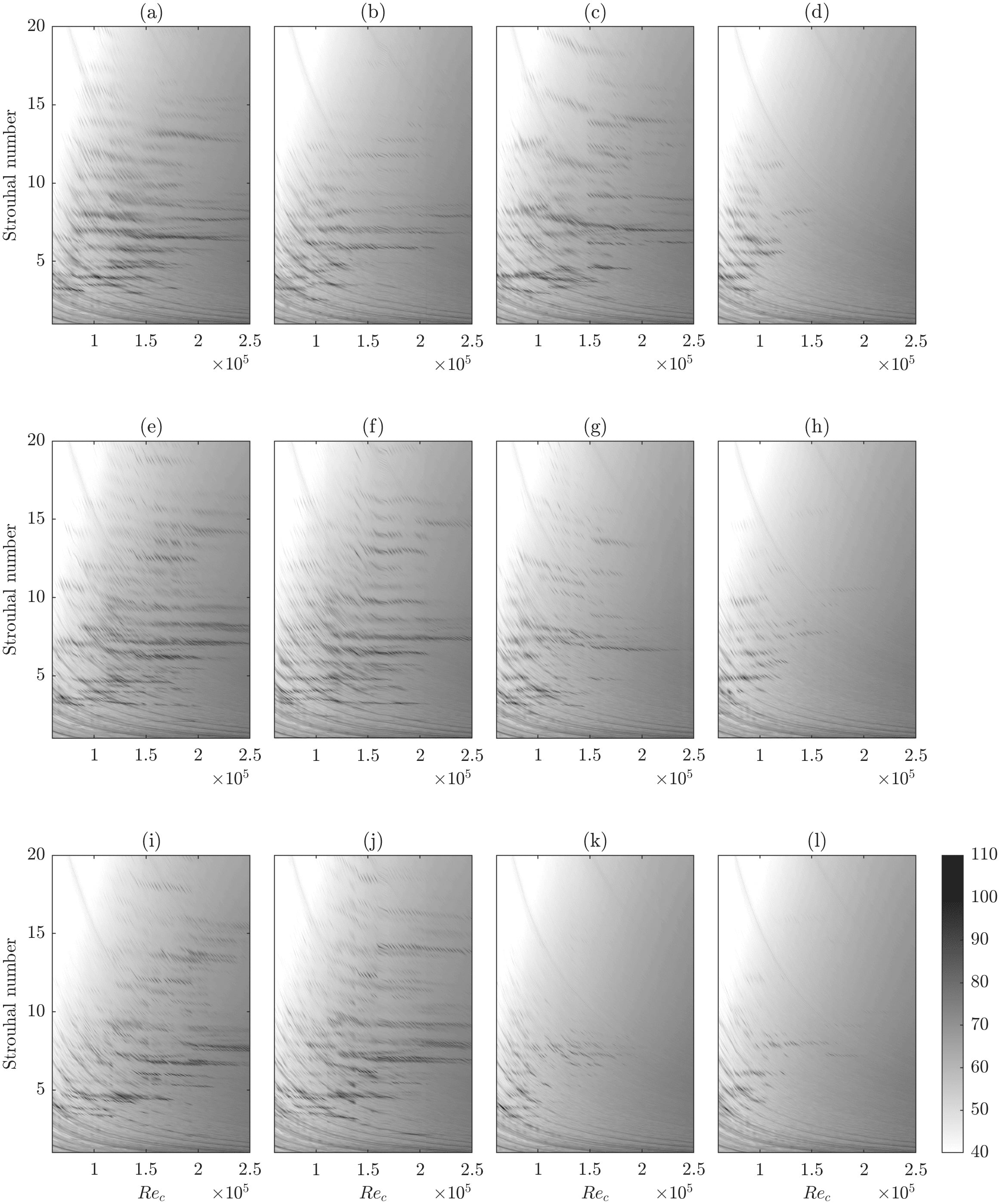}
        \caption{PSD (in dB/St) for $\alpha$ = 0, 1, and 2 degrees (frames from top to bottom), and cases: (a,e,i) smooth surface, (b,f,j) roughness on suction side, (c,g,k) roughness on pressure side, and (d,h,l) roughness on both sides (frames from left to right).}
        \label{fig:map_a012}
        \end{center}
    \end{figure}
    \begin{figure}
        \begin{center}
        \includegraphics[width=13cm]{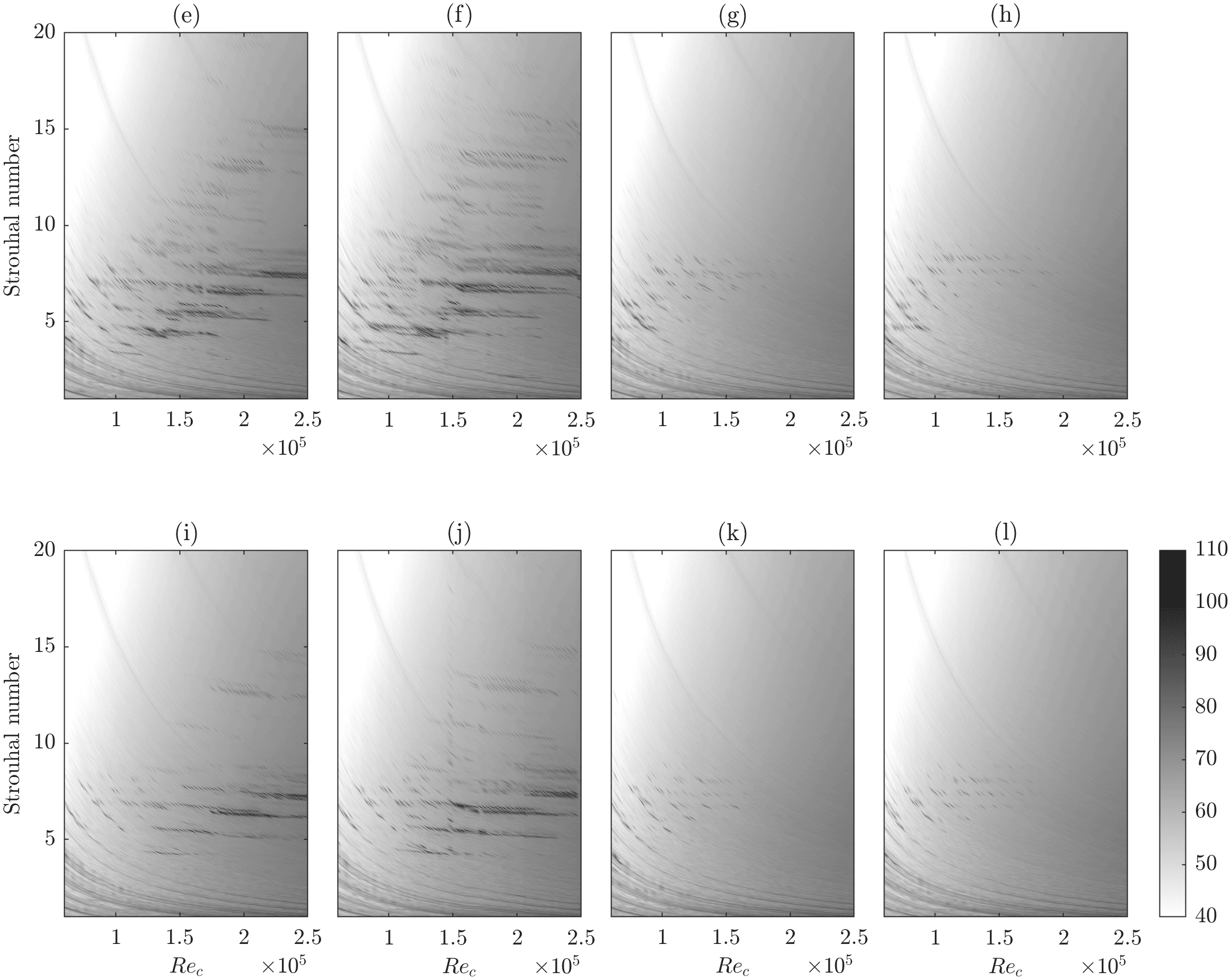}
        \caption{PSD (in dB/St) for $\alpha = 3$, and 4 degrees (top and bottom frames, respectively), and cases: (a,e) smooth surface, (b,f) roughness on suction side, (c,g) roughness on pressure side, and (d,h) roughness on both sides (frames from left to right, respectively).}
        \label{fig:map_a34}
        \end{center}
    \end{figure}
Figure \ref{fig:map_a012} shows the results of smooth and rough case for angle of attack $\alpha =$ $0$, $1$, and $2$ degrees (frames from top to bottom), the baseline and roughness configurations (frames left to right). Regarding $\alpha = 0$ degrees (top frames, from a to d), the differences between the suction side, figure \ref{fig:map_a012} (b),  and pressure side, figure \ref{fig:map_a012} (c), for the angle of attack of $0$ degrees are related to experimental uncertainties (model, setup and measurements). Some differences arise on the maps if the roughness elements are compared with tripping devices force the transition. The spectra with tripping devices of previous experiments (see \citealt{probsting2015regimes}) suppress all tonal noise, leaving only a trace of broadband noise at low $Re_c$. In the two cases with roughness elements, only on one side (b, c), there are clear reductions in the measured tones. The last case with roughness elements on both sides, figure \ref{fig:map_a012} (d), shows an improvement, especially for moderate Reynolds numbers. At low $Re_c$, some peaks are observed in the spectra, but they decrease at $Re_c = 1.5\times 10^5$. In the case with roughness elements on both sides, figure \ref{fig:map_a012} (d), the main peaks in the spectra disappear for $Re_c > 1.3 \times 10^5$. Some tones are still present for $Re_c$ lower than $1.3 \times 10^5$, but with weaker side tones. These results suggest that roughness elements are able to significantly affect K-H waves over the airfoil, which are the dominant instabilities that contributes with the radiation of the tonal noise for the separated regions \citep{nguyen2021numerical}. This will be further explorer in the companion paper \cite{Yuan2024airfoil}.

Figures \ref{fig:map_a012} (e-l) and \ref{fig:map_a34} show the results from angles of attack $\alpha = 1$ to $4$ degrees. For low angles of attack, the smooth case in figures \ref{fig:map_a012} (e,i) shows several tonal peaks, which is consistent with a previous study \citep{probsting2015regimes}. Roughness elements on the suction side in figures \ref{fig:map_a012} (f,j) do not provide significant effects. The case with roughness elements on the pressure side in figures \ref{fig:map_a012} (g,k) shows reduction of tonal noise for the entire Reynolds number range, especially for lower $Re_c$. The case with roughness elements on both sides (see figures \ref{fig:map_a012} (h,l)) has more significant tone attenuation than single-sided elements. For $Re_c > 1\times10^5$ and $\alpha = 1$ degree, the peaks of tonal noise are suppressed. For $\alpha = 2$ degrees, just weaker tones are visible.

For $\alpha =$ 3, and 4 degrees, figures \ref{fig:map_a34} (a,e) show several tonal noise peaks for the smooth case, which is consistent with a previous study \citep{probsting2015regimes}. At both angles of attack the roughness elements on the suction side (figures \ref{fig:map_a34} (b,f)) do not provide significant effects, similar to what is found for lower angles of attack. On the other hand, the presence of roughness elements on the pressure side, figures \ref{fig:map_a34} (c,g), have strong effects on tonal peaks. Most tonal noise at $Re_c > 1\times 10^5$ and $\alpha = 3$, and 4 degrees was suppressed. Roughness elements on both sides, figures \ref{fig:map_a34} (d,h), have also similiar suppression effects on tonal noise, highlighting that as the angle of attack is increased the feedback mechanism leading to tonal noise is dominated by the airfoil pressure side; accordingly, roughness elements at the pressure side are sufficient to greatly attenuate tones.

We observed some differences between our results, using the roughness elements, and those reported in the literature, using boundary-layer tripping devices. \cite{probsting2015regimes} reported that at low Reynolds numbers, the suction side dominates the tonal noise mechanism. When this side was tripped to force the boundary-layer transition, the tonal noise was suppressed. However, in our experiments, the suppression of the tonal noise was not reached with the roughness elements on the suction side for low Reynolds number and low angles of attack $\alpha = 2$, and 4 degrees. However, we observed a decrease and even suppression of tonal noise through roughness elements on the pressure side. Further, the roughness elements on the pressure side had a similar effect also at moderate Reynolds numbers, suppressing the tonal noise.
\section{Conclusion} 
\label{conclusions}

In this paper we propose the use roughness elements, in the shape of cylinders, to generate streaks around airfoils via the lift-up effect, aiming at an attenuation of tonal noise emitted by airfoils at low and moderate Reynolds numbers. The streaks are generated so as to affect the boundary layers and laminar separation bubbles around the airfoil, such that three-dimensional effects may have a stabilizing role in the Kelvin-Helmholtz mechanism \citep{marant2018influence}. Oil flow visualization experiments were performed in a NACA0012 airfoil considering smooth and rough surfaces. The baseline (smooth surface) configuration present a separation bubble on the airfoil suction side followed by a region of reattachment. On the airfoil pressure side, a separation bubble could not be observed for angles of attack different than zero. The oil flow visualization of the surface with roughness elements suggests the generation of high-speed streaks behind the cylinders, while in between cylinders, a region with low-speed streaks appears, consistent with previous experiments with laminar \citep{fransson2004experimental} and turbulent boundary layers \citep{pujals2010drag}. The effect of the streaks is greater when the roughness elements are positioned upstream of the separation bubble. On the other hand, when the roughness elements are positioned in the separation bubble region or downstream of the separation bubble, the generation of streaks is not observed. We have observed a disruption of the region of reverse flow owing to the separation bubble by the streaks on the rough surface. As the angle of attack and Reynolds number increase, the effects of roughness elements on the generation of streaks are stronger on the pressure side, while on the suction side, the roughness effect is reduced, which is likely due to the upstream movement of the separation bubble rendering the roughness elements less effective.

Detailed acoustic measurements were carried out in order to explore the effect of roughness elements, and associated streaks, on tonal noise. In some cases, the Strouhal number of the main tones is affected, highlighting that attenuation is not simply related to transition to turbulence, but to a modification of the underlying feedback loop mechanism of tonal noise. The specific effects vary as a function of angle of attack and Reynolds number, but we observe significant attenuation of tonal noise, and in some cases a complete suppression of tones. For lows angle of attack ($\alpha$ up to 2 degrees), greater attenuations of tonal noise were obtained with roughness elements at both sides of the airfoil, indicating that the feedback loop mechanisms at both sides of the airfoil are related to tonal noise, thus requiring the induction of streaks at both surfaces in order to significantly reduce tones. On the other hand, for higher angles of attack ($\alpha = 3$, and $4$ degrees), roughness elements at the pressure side alone are sufficient to attenuate most of the tonal noise.

Overall, the introduction of roughness elements was shown here as an interesting technique to modify laminar separation bubbles, which become three-dimensional as streaks are induced, as expected from earlier works \citep{karp2020optimal}. The associated three-dimensional shear layers is expected to lead to a weaker Kelvin-Helmholtz instability \citep{marant2018influence}. Although we were not able to assess these affects in such detail in the present experiments due to the complexity of velocity measurements in our setup, our companion work \cite{Yuan2024airfoil} employs large-eddy simulations of the present configurations in order to explore flow fields in more detail, confirming that the effects suggested here indeed occur. Thus, the present works provide a novel technique to reduce airfoil tonal noise, with indications for design: roughness elements capable of generating streaks, as explored in earlier works \citep{fransson2004experimental} are expected to significantly attenuate tonal noise.

\section*{Acknowledgement}
The authors acknowledge the support provided by the staff of Laboratory Prof. Kwei Lien Feng, their assistance was of great importance for the present study.

El\'ias Alva was supported from CAPES, under Grants No. 629733/2021-00 and 829471/2023-00, and from Funda\c c\~ao de Amparo à Pesquisa do Estado de S\~ao Paulo, FAPESP, under the Grant No. 2023/12528-1. Zhenyang Yuan's work was supported by the Swedish Research Council, under Grant 2020-04084.
The authors also acknowledge the financial support received from FAPESP, Grant No. 2019/27655-3, Conselho Nacional de Desenvolvimento Cient\'ifico e Tecnol\'ogico, CNPq, under Grants No. 313225/2020-6 and 407618/2022-8, and Financiadora de Estudos e Projetos, FINEP, via Helices, aerofoLios e asas Silenciosas (HELLAS) project under Grant No. 1307/22.

\bibliographystyle{jfm}
\bibliography{bibpartI}

\end{document}